%% file: main.tex
\documentclass[runningheads]{llncs}
\usepackage[T1]{fontenc}
\usepackage{soul}
\usepackage{packages}

\apptocmd{\normalsize}{\setlength{\tabcolsep}{3pt}}{}{} 

\title{Does Model Size Matter? A Comparison\\ of Small and Large Language Models \\for Requirements Classification}
\titlerunning{Does Model Size Matter?}
\author{
Mohammad Amin Zadenoori\inst{1}\orcidID{0000-0003-4591-153X} \and
Vincenzo De Martino\inst{2,}\inst{3}\orcidID{0000-0003-1485-4560} \and
Jacek Dąbrowski\inst{4}\orcidID{0000-0003-3392-0690} \and \\
Xavier Franch\inst{3}\orcidID{0000-0001-9733-8830} \and
Alessio Ferrari\inst{5}\orcidID{0000-0002-0636-5663}
}

\authorrunning{Zadenoori et al.}

\institute{
University of Padova, Italy \\
\email{amin.zadenoori@unipd.it} \and
Software Engineering (SeSa) Lab, University of Salerno, Italy \\
\email{vdemartino@unisa.it} \and
Universitat Politècnica de Catalunya, Spain \\
\email{\{vincenzo.de.martino, xavier.franch\}@upc.edu} 
\and
Lero, the Research Ireland Centre for Software, University of Limerick, Ireland \\
\email{jacek.dabrowski@lero.ie} \and
University College Dublin (UCD), Ireland \\
\email{alessio.ferrari@ucd.ie} 
}

\date{August 2025}

\begin{document}

\maketitle
\begin{abstract}
\textbf{[Context and motivation]} Large language models (LLMs) show notable results in natural language processing (NLP) tasks for requirements engineering (RE). However, their use is compromised by high computational cost, data-sharing risks, and dependence on external services. In contrast, small language models (SLMs) offer a lightweight, locally deployable alternative. \textbf{[Question/problem]} It remains unclear how well SLMs perform compared to LLMs in RE tasks in terms of accuracy. \textbf{[Results]} Our preliminary study compares eight models, including three LLMs and five SLMs, on requirements-classification tasks using the PROMISE, PROMISE Reclass, and SecReq datasets. 
Our results show that although LLMs achieve an average F1 score of 2\% higher than SLMs, this difference is not statistically significant. SLMs almost reach LLMs performance across all datasets and even outperform them in recall on the PROMISE Reclass dataset, despite being up to 300 times smaller. We also found that dataset characteristics play a more significant role in performance than model size. \textbf{[Contribution]} Our study contributes with evidence that SLMs are a valid alternative to LLMs for requirements classification, offering advantages in privacy, cost, and local deployability.

\end{abstract}

\input{Sections/introduction}
\input{Sections/EmpericalStudy}

\input{Sections/Results}

\input{Sections/researchplan}

\begin{credits}
    
\vspace{-0.3cm}
\subsubsection*{Acknowledgements} 
This work was supported in part by project EOSS6-0000000644 from the Chan Zuckerberg Initiative. Grok-4 (\href{Grok}{https://x.ai/grok})  was used to improve the readability, and the authors remain fully responsible for the final content.
\vspace{-0.3cm}
\subsubsection{\discintname}
The authors have no competing interests to declare that are relevant to the content of this article. 

\end{credits}

\bibliographystyle{splncs04}
\bibliography{bibliography}
\end{document}

%% file: Sections/introduction.tex
\section{Introduction}
\label{lab:introduction}

Requirements classification is a critical task in requirements engineering (RE). It involves categorizing requirements into types such as functional/non-functional, or more fine-grained classes (e.g., security, performance)~\cite{10.1145/3444689}. This task supports other RE activities, including requirements management and traceability~\cite{alhoshan2023zero}. 
However, as projects grow larger, the number of requirements can reach thousands, making manual classification labor-intensive and error-prone. This has motivated research on automating requirements classification using natural language processing (NLP) techniques~\cite{10.1145/3444689,Dabrowski2022}.
Recent advances in large language models (LLMs), such as GPT and Claude, have transformed NLP, achieving state-of-the-art results in text understanding, classification, and generation~\cite{zadenoori2025largelanguagemodelsllms}. Consequently, the RE community is increasingly exploring their use for RE tasks, particularly automated requirements classification~\cite{alhoshan2025effectivegenerativelargelanguage}.

Despite their power, LLMs raise concerns about privacy, security, and reproducibility. They are typically closed-source and cloud-hosted, which increases the risk of data exposure, as company requirements constitute confidential assets~\cite{ferrari2017natural} and the cloud operates as an external service. Their proprietary nature also limits researchers’ ability to adapt models for RE-specific needs and reduces their practicality in industrial settings. Open small language models (SLMs) offer a viable alternative for local execution on private machines or servers, enabling secure processing of sensitive data. Local deployment also lowers costs and enables easier customization and fine-tuning \cite{wang2024comprehensive}. Furthermore, they offer advantages in terms of energy consumption \cite{de2025greenprompt}. 
However, despite their advantages, the performance comparison between SLMs and LLMs in RE tasks remains unexplored.


This study addresses that gap through a systematic comparison of eight language models. We include five SLMs with 7-8 billion parameters (\textit{Qwen2-7B}, \textit{Falcon-7B}, \textit{Granite-3.2-8B}, \textit{Ministral-8B}, and \textit{Llama-3-8B}) and three LLMs (\textit{GPT-5}, \textit{Grok-4}, and \textit{Claude-4}), with 1-2 trillion parameters. The models were selected based on the top-performing LLMs in the Hugging Face Open {\it LLM Leaderboard}\footnote{\texttt{https://huggingface.co/spaces/open-llm-leaderboard}} to ensure a fair comparison.
We evaluated all models on three public datasets: \textit{PROMISE} \cite{Cleland-Huang2007Promise}, a re-classification of PROMISE that we call \textit{PROMISE Reclass} \cite{inproceedingsPromisereclass}, and \textit{SecReq} \cite{inproceedingssecReq}. Our results show that SLMs perform comparably to LLMs, although the latter are 100 to 300 times larger in terms of parameters. Among all LLMs, \textit{Claude-4} consistently achieves the highest F1 scores, including 0.81 on PROMISE, 0.80 on PROMISE Reclass, and 0.89 on SecReq. The best-performing SLM, \textit{Llama-3-8B}, reached 0.76, 0.78, and 0.88 on the same tasks, respectively, yielding an average F1 margin of 0.02 in favor of the LLMs. Our results suggest that model size has a limited effect on classification accuracy,  while SLMs offer a promising balance between performance, privacy, and resource efficiency. Furthermore, we show that performance highly depends on the dataset used for classification.
\textbf{Data Availability Statement.} To promote transparency and reproducibility, we make our replication package publicly available~\cite{appendix}.

%% file: Sections/EmpericalStudy.tex
\vspace{-0.3cm}
\section{Study Design}
\label{sec:Empdesign}
\vspace{-0.3cm}
Our study aims to address the following research question \textbf{(RQ):} 
\rqbox{
\textit{What is the performance difference between SLMs and LLMs in requirements classification tasks?}
}

To address our RQ, we designed a reproducible pipeline, described in Table~\ref{tab:study_design}. It should be noted that, although some open-source LLMs now approach or exceed 100B parameters (e.g., Falcon-180B, Mixtral-8x22B) and can be executed on personal servers, \emph{LLMs} in this study refer exclusively to proprietary models maintained by private companies and accessed via commercial APIs (e.g., GPT-5, Claude-4, Grok-4). In contrast, \emph{SLMs} denote openly accessible models (e.g., Qwen2-7B, Falcon-7B) that support local deployment. The evaluated SLMs range from 7B to 8B parameters, whereas LLMs can reach trillions, although their precise sizes remain undisclosed.

\begin{table}[H]
\centering
\scriptsize
\caption{Study Design Overview Following the PRIMES Framework \cite{de2025framework}}
\label{tab:study_design}
\begin{tabularx}{\textwidth}{|p{2cm}|X|}
\hline
\textbf{Component} & \textbf{Description} \\
\hline
\textbf{Datasets} & 
 \textbf{PROMISE}~\cite{Cleland-Huang2007Promise}: Binary classification of Functional Requirements (FR) vs. Non-Functional Requirements (NFR). Composed of 625 requirements (255 FR,
    370 NFR).
    
\textbf{PROMISE Reclass}~\cite{inproceedingsPromisereclass}:  Two binary subtasks: FR vs. NFR and QR vs. Non-QR. Each requirement has two labels. Composed of 625 requirements (310 FR, 230 FR only; 382 QR, 302 QR only).

\textbf{SecReq}~\cite{inproceedingssecReq}:Binary classification of Security-related (Sec) vs. Non-Security (NSec) requirements. Composed of 510 requirements (187 Sec, 323 NSec).
 \\
\hline
\textbf{Models} & 
\textbf{SLMs:} Qwen2-7B-Instruct, Falcon-7B-Instruct, Granite-3.2-8B-Instruct, Ministral-8B-Instruct-2410, Meta-Llama-3-8B-Instruct.

    \textbf{LLMs:} GPT-5, xAI Grok-4, Claude-4.
\\
\hline
\textbf{Prompting Strategy} & \textbf{Chain-of-Thought} (CoT) plus \textbf{Few-Shot} prompting with four examples per class, which—based on our prior work~\cite{10.1007/978-3-031-88531-0_15}—emerged as the most effective among all evaluated prompting strategies. Definitions of categories are based on expert-defined RE definitions~\cite{Olsson2022,10.1007/978-3-642-14192-8_15}. \\
\hline
\textbf{Parameters} & Default parameters; temperature fixed at 0 for deterministic outputs. \\
\hline
\textbf{Hardware and software} & SLMs executed on Linux 6.14 server with Intel i9-13900K CPU, 128~GB RAM, NVIDIA RTX 4090 GPU, Python 3.12. LLMs accessed via commercial APIs. \\
\hline
\textbf{Runs} & Each model evaluated on all datasets, with three executions per task. Majority voting (2/3) is used for the final label. Metrics computed via macro-averaging. \\
\hline
\textbf{Statistical test} & Scheirer-Ray-Hare test, a non-parametric equivalent of two-way ANOVA. \\
\hline
\end{tabularx}
\end{table}

%% file: Sections/Results.tex
\section{Preliminary Results and Discussion}
\label{sec:results}
This section presents the preliminary results of our investigation, analyzing the performance of the evaluated models across the selected datasets. Table~\ref{tab:combined_results} reports classification performance per dataset, including average metrics, considering precision (P), recall (R), and F1. Our evaluation employs these standard metrics without assuming any specific trade-off between P and R, as their relative importance varies across application contexts. 


\begin{table}[t]
\centering
\scriptsize
\caption{Results across all datasets using CoT $\cup$ Few-shot.}
\label{tab:combined_results}
\begin{tabular}{l|ccc|ccc|ccc|}
\cline{2-10}
 & \multicolumn{3}{c|}{\textbf{PROMISE}} & \multicolumn{3}{c|}{\textbf{PROMISE Reclass}} & \multicolumn{3}{c|}{\textbf{SecReq}} \\ 
\cline{2-4} \cline{5-7} \cline{8-10}
 & \textbf{P} & \textbf{R} & \textbf{F1} & \textbf{P} & \textbf{R} & \textbf{F1} & \textbf{P} & \textbf{R} & \textbf{F1} \\ 
\hline
\multicolumn{1}{|l|}{Qwen2-7B} & 0.86 & 0.80 & 0.83 & 0.55 & \textbf{0.96} & 0.70 & 0.83 & 0.90 & 0.86 \\ 
\multicolumn{1}{|l|}{Falcon3-7B} & 0.86 & 0.80 & 0.83 & 0.58 & \textbf{0.96} & 0.72 & 0.79 & 0.90 & 0.84 \\ 
\multicolumn{1}{|l|}{Granite-3.2} & 0.84 & 0.82 & 0.83 & 0.62 & 0.87 & 0.72 & 0.84 & 0.90 & 0.87 \\ 
\multicolumn{1}{|l|}{Ministral-8B} & 0.84 & 0.77 & 0.80 & 0.64 & 0.89 & 0.75 & 0.83 & 0.88 & 0.85 \\ 
\multicolumn{1}{|l|}{Llama-3-8B} & 0.84 & 0.77 & 0.80 & 0.70 & 0.87 & 0.78 & 0.86 & 0.91 & 0.88 \\
\hline
\multicolumn{1}{|l|}{\textbf{SLMs (Mean)}} & 0.85 & 0.79 & 0.82 & 0.62 & 0.91 & 0.73 & 0.83 & 0.90 & 0.86 \\ 
\hline
\multicolumn{1}{|l|}{Grok-4} & \textbf{0.88} & \textbf{0.83} & \textbf{0.85} & 0.60 & 0.82 & 0.69 & 0.84 & 0.89 & 0.86 \\ 
\multicolumn{1}{|l|}{GPT-5o} & 0.85 & 0.81 & 0.83 & 0.68 & 0.88 & 0.77 & 0.85 & 0.90 & 0.87 \\ 
\multicolumn{1}{|l|}{Claude-4} & 0.85 & 0.80 & 0.82 & \textbf{0.72} & 0.90 & \textbf{0.80} & \textbf{0.87} & \textbf{0.92} & \textbf{0.89} \\
\hline
\multicolumn{1}{|l|}{\textbf{LLMs (Mean)}} & 0.86 & 0.81 & 0.83 & 0.67 & 0.87 & 0.75 & 0.85 & 0.90 & 0.88 \\ 
\hline
\end{tabular}
\end{table}

\smallskip
\noindent
\textbf{Overall Performance.} Across the evaluated classification tasks, LLMs demonstrate a consistent, albeit subtle, performance advantage over SLMs. However, a detailed examination reveals a more complex picture where SLMs are highly competitive and even excel in specific areas. While LLMs lead in average F1 score in all datasets, the performance gap is not uniform. In fact, in all datasets and metrics, we observe that there is always at least one SLM behaving better than one LLM.  
In the PROMISE dataset, the leading SLMs (Qwen2-7B, Falcon3-7B, and Granite-3.2) achieve an F1 score of 0.83, effectively tying with the best LLM, Grok-4 (0.85). More notably, in the PROMISE Reclass dataset, SLMs demonstrate a remarkable advantage in Recall, with Qwen2-7B and Falcon3-7B reaching 0.96—significantly higher than any LLM. This indicates that SLMs are exceptionally effective at identifying all relevant instances within this specific context, minimizing false negatives. Furthermore, the SLM Ministral-8B achieves a Precision of 0.64 in PROMISE Reclass, outperforming the LLM Grok-4 (0.60), suggesting greater reliability in its positive predictions for that dataset. 

Even in the SecReq dataset, the performance difference is minimal, with the top SLM, Llama-3-8B (F1: 0.88), closely trailing the top LLM, Claude-4 (F1: 0.89). These nuanced results indicate that while LLMs possess a slight edge in balanced overall performance (F1), SLMs are not merely competitive; they can peak higher than LLMs in isolated but crucial metrics like Recall and Precision. This suggests that SLMs have specialized strengths that make them highly capable for specific classification scenarios where minimizing certain types of errors is paramount. To summarize our performance findings in each dataset:

\begin{itemize}
  \item\textbf{PROMISE:} Grok-4 (LLM) shows best behaviour for all metrics, but interestingly three SLMs behave similarly than the other two LLMs. As result, averages are quite close among both types.
  \item \textbf{PROMISE Reclass:} We notice remarkable variations of different sign in P and R, and consequently a drop in F1. LLMs behave better in P and F1 (with Claude-4 as best) but SLMs exhibit higher R. 
   \item \textbf{SecReq:} Claude-4 also dominates in this dataset, remarkable in all metrics. Behaviour of all other LLMs and SLMs is quite close to each other, with an overall minor dominance on the side of LLMs.

       
\end{itemize}
\smallskip
\noindent
\textbf{Tests and Hypothesis.} We next formulate and test our hypothesis. Although our RQ focuses on performance differences, here we also consider the \textit{dataset} variable, as this could be a factor influencing the results. The null hypotheses are: (i) $H_{0A}$ model type has no effect on performance, (ii) $H_{0B}$ dataset has no effect on performance, and (iii) $H_{0C}$ there is no interaction between model type and dataset. Since the sample distributions did not meet the assumptions of normality, as assessed by the \textbf{Shapiro-Wilk test}, we used the \textbf{Scheirer-Ray-Hare test}, a non-parametric equivalent of two-way ANOVA (see results in Table \ref{tab:srh_results}). The post-hoc tests are follow-up analyses that break down the overall effects reported in \ref{tab:srh_results}, specifying whether differences between model types or datasets are statistically significant within individual comparisons. Post-hoc analyses were conducted using pairwise Mann-Whitney U tests with Bonferroni correction.

\begin{table}[t]
\centering
\scriptsize
\caption{Scheirer-Ray-Hare Test Results for Model Performance (F1 Score). Effect sizes: $\eta^2_H$ = 0.01 (small), 0.06 (medium), 0.14 (large).}
\label{tab:srh_results}
\begin{tabular}{lllll}
\toprule
\textbf{Hp} & \textbf{Variable} & \textbf{Effect Size ($\eta^2_H$)} & \textbf{p-value} & \textbf{Reject/Fail to Reject} \\
\midrule
$H_{0A}$ & Model Type & 0.04 & 0.296 & Fail to Reject $H_{0A}$\\
$H_{0B}$ & Dataset & 0.63 & \textbf{<0.001} & \textbf{Reject} $H_{0B}$\\
$H_{0C}$ & Model Type $\times$ Dataset & 0.001 & 0.790 & Fail to Reject $H_{0C}$\\
\bottomrule
\end{tabular}

\end{table}

\noindent
\textbf{Post-hoc Model Type Comparisons within each Dataset:}
\begin{itemize}
    \item PROMISE: U = 7.0, p = 0.134 (Not Significant)
    \item PROMISE\_Reclass: U = 11.0, p = 0.764 (Not Significant)
    \item SecReq: U = 7.0, p = 0.291 (Not Significant)
\end{itemize}

\noindent
\textbf{Post-hoc Pairwise Dataset Comparisons (Bonferroni-corrected):}
\begin{itemize}
    \item PROMISE\_Reclass vs. PROMISE: U = 6.50, p = 0.0084 (Significant)
    \item PROMISE\_Reclass vs. SecReq: U = 0.00, p = 0.0009 (Significant)
    \item PROMISE vs. SecReq: U = 3.50, p = 0.0031 (Significant)
\end{itemize}

\smallskip
\noindent
\textbf{Interpretation of Findings.} Based on the results, the following conclusions can be derived.

\begin{enumerate}
    \item \textbf{Model Type Effect (SLM vs. LLM):} 
    The analysis confirms no statistically significant main effect of model type on F1 score (H = 1.09, p = 0.296, and small effect size $\eta^2_H = 0.04$). Despite descriptive statistics showing slightly higher mean F1 scores for LLMs (0.818) compared to SLMs (0.793), this difference is not large enough to be considered statistically significant. 

    \item \textbf{Dataset Effect:}
    We found a highly significant main effect of the dataset on F1 score (H = 17.53, p < 0.001, and large effect size $\eta^2_H = 0.63$). Post-hoc tests revealed that all pairwise differences between datasets are statistically significant, with a clear performance hierarchy: models performed worst on \textbf{PROMISE\_Reclass} (Median F1 = 0.730), better on \textbf{PROMISE} (Median F1 = 0.805), and best on \textbf{SecReq} (Median F1 = 0.865).

    \item \textbf{Interaction Effect (Model Type $\times$ Dataset):}
    Critically, the \textbf{interaction between Model Type and Dataset was not significant} (H = 0.47, p = 0.790, and small effect size $\eta^2_H = 0.001$). Since the interaction is not significant, the effect of Dataset observed above is consistent across models---\textbf{all  models are affected by datasets in roughly the same way}. This is further supported by the post-hoc tests within each dataset, which all failed to show a significant difference between SLMs and LLMs on any individual dataset.
\end{enumerate}

\smallskip
\noindent
It should be noted that, while the analysis found no statistically significant performance difference between SLMs and LLMs ($p = 0.296$), the consistent directional advantage for LLM type (represented by the mean of all models) across all datasets 
might indicate a high probability of Type II error (i.e., we fail to reject $H_{0A}$, although there is a difference between the performance of SLMs and LLMs). The limited sample size of eight models substantially reduces statistical power, likely obscuring genuine performance differences. Nevertheless, even if the difference is confirmed to be significant by future studies, we argue that a loss of 2\% can be considered acceptable, given the advantages of SLMs in data privacy and resource efficiency.
\rqanswer{LLMs show a slight edge in performance compared to SLMs, leading by ~2\% in F1. The difference is, however, not statistically significant. Furthermore, when considering the recall metric on specific datasets, SLMs tend to outperform LLMs.}

%% file: Sections/researchplan.tex
\smallskip
\noindent

\smallskip
\noindent
\textbf{Threats to Validity.} \textit{Construct.} We have used traditional metrics for evaluation, and not accounted for $F_\beta$. Further experiments are required to identify the appropriate $\beta$ value for our context. These were not conducted, given the preliminary nature of the study. \textit{Internal.} The same prompt was applied across all models to ensure consistency; however, relying on a single prompt, even if informed by prior studies~\cite{10.1007/978-3-031-88531-0_15}, may affect internal validity since model performance could depend on prompt phrasing~\cite{alhoshan2025effectivegenerativelargelanguage}. Internal validity is also affected by the variability of model output. To address this, we repeated each run three times, and performed majority voting. There is also a risk of data leakage, as some datasets may have been included in the pre-training corpora of the evaluated LLMs, potentially inflating their performance. This could not be entirely mitigated. \textit{External.} We have considered the binary classification task, so our results might not generalise to other RE tasks. On the other hand, we have used three different datasets, which increase the generalisability of our results, at least for the task at hand \textit{Conclusion.} We performed appropriate statistical tests to check our conclusions. However, the limited sample size might have led to Type II errors. This is justified by the preliminary nature of the experiments.  To prevent multiple comparison problems arising from different dependent variables, we used only the F1 metric for statistical analysis, which might not account for all performance nuances. To address this, we have complemented the results with observations on precision and recall for specific cases, and reported the values in Table~\ref{tab:combined_results}.

\vspace{-0.3cm}
\section{Conclusion and Roadmap}
\vspace{-0.3cm}
Our study compared LLMs and SLMs performance in requirements classification, showing that the formed gives only marginal F1 improvement (2\%) over the latter. This suggests that, at least for this RE tasks, companies can profit for the advantages of SLMs in terms of data privacy and resource demands. The following research avenues will be explored in our future research. 

\textit{Explainability.} While LLMs and SLMs have shown strong and comparable performance in requirements classification, such tasks do not fully leverage their generative capabilities: classification alone captures the outcome of reasoning but not the reasoning process itself. To exploit the full potential of these models, we should move beyond simple label prediction and focus on generating \textit{explanations} that justify classifications. Explanations are particularly important in RE, as they enable practitioners to make informed decisions based on the rationale behind classifications rather than on predictions alone. Understanding why a requirement is categorized in a certain way supports confidence in automated analyses, as well as more informed downstream tasks, e.g., traceability, and auditing. 

\textit{Different RE Tasks and Hybrid Pipelines.} We plan to extend the evaluation to additional RE tasks such as traceability, model generation, and requirements elicitation, which have already seen interest in LLM for RE~\cite{zadenoori2025largelanguagemodelsllms}. Special attention will be given to tasks that require the generation of artifacts, as these are the ones where generative models can offer a true competitive advantage with respect to more traditional machine learning (ML) solutions. Experimenting with SLMs and LLMs in different tasks will potentially enable the identification of hybrid pipelines for RE workflows, where each model type is used for the task or phase where it brings the best advantage, e.g., traditional ML models for classification, SLMs for explanation, and LLMs for supporting analyst interaction and iterative refinement through conversational assistance.

\textit{Energy Consumption.} Sustainability of AI computation is a major concern, as LLMs can have substantial environmental impact---although still lower than manual labour~\cite{ren2024reconciling}. Further research is needed to quantify the energy footprint of RE-specific tasks, comparing SLMs and LLMs. This includes evaluating trade-offs between model performance and energy efficiency, especially in local vs. cloud-based deployments of SLMs and LLMs \cite{de2025greenprompt}. 

\textit{Execution Speed.} Systematically studying speed differences between SLMs and LLMs is also an important avenue to consider, as RE activities need to be accurate but also fast, to cope with the ever-changing nature of stakeholder demands. In our study, Claude-4 is the fastest on PROMISE Reclass (300s), Grok-4 leads on PROMISE (162s), and GPT-5o is the quickest on SecReq (138s), while the computation time of SLMs is on average 400s. This gap is expected, given that proprietary LLMs are deployed on high-performance cloud infrastructure, whereas the SLMs in this study were executed on a single local server given in Table \ref{tab:study_design}. 
Although the execution speed is specific to our set-up and can widely vary, we believe that these numbers give an indication of the order of magnitude
of the time difference that a small-medium company—--which can be considered
similar in terms of resources to a research lab such as ours—--can expect when
using local SLMs. Further studies are needed to understand what speed range is practically acceptable for practitioners and how to optimise execution efficiency.


\textit{Tool Support.} 
Deployment of LLMs in companies is a major pain point, and comparison between SLM/LLM solutions can be cumbersome. As future work, we envision the development of a tool that enables practitioners to flexibly select between locally hosted SLMs and API-based LLMs for requirements classification. 
The tool will adapt model selection and prompt strategies according to specific constraints, such as data privacy, computational resources, and task complexity, thus supporting context-aware and efficient RE workflows.

%% file: main.bbl
\begin{thebibliography}{10}
\providecommand{\url}[1]{\texttt{#1}}
\providecommand{\urlprefix}{URL }
\providecommand{\doi}[1]{https://doi.org/#1}

\bibitem{alhoshan2023zero}
Alhoshan, W., Ferrari, A., Zhao, L.: Zero-shot learning for requirements
  classification: An exploratory study. Information and Software Technology
  \textbf{159},  107202 (2023)

\bibitem{alhoshan2025effectivegenerativelargelanguage}
Alhoshan, W., Ferrari, A., Zhao, L.: How effective are generative large
  language models in performing requirements classification? (2025),
  \url{https://arxiv.org/abs/2504.16768}

\bibitem{Cleland-Huang2007Promise}
Cleland-Huang, J., Settimi, R., Zou, X., Solc, P.: Automated classification of
  non-functional requirements. Requirements Engineering  \textbf{12}(2),
  103--120 (Apr 2007). \doi{10.1007/s00766-007-0045-1}

\bibitem{inproceedingsPromisereclass}
Dalpiaz, F., Dell'Anna, D., Aydemir, F., Cevikol, S.: Requirements
  classification with interpretable machine learning and dependency parsing.
  pp. 142--152 (09 2019). \doi{10.1109/RE.2019.00025}

\bibitem{de2025framework}
De~Martino, V., Casta{\~n}o, J., Palomba, F., Franch, X.,
  Mart{\'\i}nez-Fern{\'a}ndez, S.: A framework for using llms for repository
  mining studies in empirical software engineering. In: 2025 IEEE/ACM
  International Workshop on Methodological Issues with Empirical Studies in
  Software Engineering (WSESE). pp. 6--11. IEEE (2025)

\bibitem{de2025greenprompt}
De~Martino, V., Zadenoori, M.A., Franch, X., Ferrari, A.: Green prompt
  engineering: Investigating the energy impact of prompt design in software
  engineering (2025)

\bibitem{Dabrowski2022}
Dąbrowski, J., Letier, E., Perini, A., Susi, A.: Analysing app reviews for
  software engineering: a systematic literature review. Empirical Software
  Engineering  \textbf{27}(2), ~43 (2022). \doi{10.1007/s10664-021-10065-7}

\bibitem{10.1007/978-3-642-14192-8_15}
Ernst, N.A., Mylopoulos, J.: On the perception of software quality requirements
  during the project lifecycle. In: Wieringa, R., Persson, A. (eds.) REFSQ. pp.
  143--157. Springer Berlin Heidelberg, Berlin, Heidelberg (2010)

\bibitem{ferrari2017natural}
Ferrari, A., Dell'Orletta, F., Esuli, A., Gervasi, V., Gnesi, S., et~al.:
  Natural language requirements processing: a {4D} vision. IEEE SOFTWARE
  \textbf{34}(6),  28--35 (2017)

\bibitem{inproceedingssecReq}
Knauss, E., Houmb, S., Schneider, K., Islam, S., Jürjens, J.: Supporting
  requirements engineers in recognising security issues. In: REFSQ 2011. pp.
  4--18. \doi{10.1007/978-3-642-19858-8_2}

\bibitem{Olsson2022}
Olsson, T., Sentilles, S., Papatheocharous, E.: A systematic literature review
  of empirical research on quality requirements. Requirements Engineering
  \textbf{27}(2),  249--271 (Jun 2022). \doi{10.1007/s00766-022-00373-9}

\bibitem{ren2024reconciling}
Ren, S., Tomlinson, B., Black, R.W., Torrance, A.W.: Reconciling the
  contrasting narratives on the environmental impact of large language models.
  Scientific Reports  \textbf{14}(1),  26310 (2024)

\bibitem{wang2024comprehensive}
Wang, F., et~al.: A comprehensive survey of small language models in the era of
  large language models: Techniques, enhancements, applications, collaboration
  with llms, and trustworthiness. preprint arXiv:2411.03350  (2024)

\bibitem{appendix}
Zadenoori, A.: Comparison of small and large language models for requirements
  classification (Oct 2025), \url{https://doi.org/10.5281/zenodo.17339105}

\bibitem{zadenoori2025largelanguagemodelsllms}
Zadenoori, M.A., Dąbrowski, J., Alhoshan, W., Zhao, L., Ferrari, A.: Large
  language models ({LLMs}) for requirements engineering ({RE}): A systematic
  literature review (2025), \url{https://arxiv.org/abs/2509.11446}

\bibitem{10.1007/978-3-031-88531-0_15}
Zadenoori, M.A., Zhao, L., Alhoshan, W., Ferrari, A.: Automatic prompt
  engineering: The case of requirements classification. In: Hess, A., Susi,
  A. (eds.) REFSQ. pp. 217--225. Springer Nature Switzerland, Cham (2025)

\bibitem{10.1145/3444689}
Zhao, L., Alhoshan, W., Ferrari, A., Letsholo, K.J., Ajagbe, M.A., Chioasca,
  E.V., Batista-Navarro, R.T.: Natural language processing for requirements
  engineering: A systematic mapping study. ACM Comput. Surv.  \textbf{54}(3)
  (Apr 2021)

\end{thebibliography}
